\documentclass[apj]{emulateapj}
\begin{document}
\def\beq{\begin{equation}}
\def\eeq{\end{equation}}
\def\bey{\begin{eqnarray}}
\def\eey{\end{eqnarray}}
\def\pc{\, {\rm pc} }
\def\kpc{\, {\rm kpc} }
\def\msun{M_\odot}
\def\sun{\odot}
\def\lsim{\mathrel{\raise.3ex\hbox{$<$\kern-.75em\lower1ex\hbox{$\sim$}}}}
\def\gsim{\mathrel{\raise.3ex\hbox{$  $\kern-.75em\lower1ex\hbox{$\sim$}}}}
\def\Msun{M_\odot}
\def\Lsun{L_\odot}
\def\lsun{L_\odot}
\def\kms{\, {\rm km \, s}^{-1} }
\def\eV{\, {\rm eV} }
\def\keV{\, {\rm keV} }
\def\dis{{\rm dis}}
\def\grad{{\bf \nabla}}
\def\HSZ{{\it HSZ}}
\title{On the Proof of Dark Matter, the Law of Gravity and the Mass of Neutrinos}
\author{
Garry W. Angus\altaffilmark{1},
HuanYuan Shan\altaffilmark{2,1},
HongSheng Zhao\altaffilmark{1,2},
Benoit Famaey\altaffilmark{3}}
\email{gwa2@st-andrews.ac.uk}
\altaffiltext{1}{
SUPA, School of Physics and Astronomy, University of St Andrews, KY16 9SS, UK
}
\altaffiltext{2}{
National Astronomical Observatories, Chinese 
Academy of Sciences, Beijing 100012, PRC
}
\altaffiltext{3}{
Institut d'Astronomie et d'Astrophysique, Universite Libre de Bruxelles, BELGIUM
}
\begin{abstract}
We develop a new method to predict the density associated with weak lensing maps of (un)relaxed clusters
in a range of theories interpolating between GR and
MOND (General Relativity and Modified Newtonian Dynamics). 
We apply it to fit the lensing map of the bullet merging
cluster 1E0657-56, in order to constrain
more robustly the nature and amount of collisionless matter in
clusters {\it beyond} the usual assumption of spherical equilibrium
(Pointecouteau \& Silk 2005) and the validity of GR on cluster scales 
(Clowe et al. 2006).  Strengthening the proposal of previous authors we
show that the bullet cluster is dominated by a collisionless -- most
probably non-baryonic -- component in GR as well as in MOND, a result
consistent with the dynamics of many X-ray clusters.  Our findings add to
the number of known pathologies for a purely baryonic MOND, including its
inability to fit the latest data from the Wilkinson Microwave Anisotropy
Probe.  A plausible resolution of all these issues and standard issues of
Cold Dark Matter with galaxy rotation curves is
the ``marriage" of MOND with ordinary hot neutrinos of 2eV.
This prediction is just within the GR-independent maximum of neutrino mass 
from current $\beta$-decay experiments, and is falsifiable by the
Karlsruhe Tritium Neutrino (KATRIN) experiment by 2009.
Issues of consistency with strong lensing arcs and the large relative velocity of the two clusters comprising the bullet cluster are also addressed.
\end{abstract}
\maketitle

The bullet interacting cluster 1E 0657-56 has recently been argued to
have produced the first completely unambiguous evidence that galaxy
clusters are shrouded in a dominant component of collisionless dark
matter (Clowe et al. 2006, hereafter C06). 
The explanation for this was that the peaks of the convergence
map are offset, without a shadow of a doubt, from 
the main observable baryonic components, i.e.,
the gas marked by the bright X-ray emission. Instead, the lensing signal peaks
at the galaxies 
(the minor observed baryonic components, about 1/7 of the total X-ray gas mass)
which is exactly where we expect any collisionless dark matter (DM) halos 
to center on.  It was also argued that modified gravity theories, including
the relativistic counterpart of MOND (Milgrom 1983, Bekenstein 2004), 
would have no way of reproducing the lensing map since the gravity 
in these theories is thought to trace the light.

There are, however, two caveats with this line of reasoning. As first cautioned
by Angus, Famaey \& Zhao  (2006, hereafter AFZ06), the features in the
lensing convergence  map in  a non-linear gravity theory do  {\it not
always} reflect  features in the underlying matter surface density in
highly non-spherical  geometries.  For example, in MOND,  the convergence can
indeed be {\it  non-zero\/} where there is no  projected matter (Zhao,
Bacon, Taylor \&  Horne 2006, Zhao \& Qin 2006). The other caveat is
whether ordinary collisonless neutrinos (which are detected particles known to have a small mass, although non-baryonic and non-photon-emitting as all leptons in the standard model are) should be given equal status as the ``known'' matter (e.g., the hadrons in gas and stars) or the stigma of ``unknown'' dark matter 
(e.g. primodial black holes, exotic cold WIMPs from SUperSYmmetry).
Indeed, ordinary neutrinos represent at least 0.1-3 times the density of 
``known'' gas and stars in a galaxy cluster
given the current experimental mass range of $N \ge 3$ flavors of neutrinos
$\sum_{i=1}^N m_{\nu,i} = 3\times (0.07\eV -2.2\eV) = 
{\Omega_\nu \over 0.004-0.125} \left({h \over 0.7}\right)^{2}$.

As for MOND, the real question is actually to examine whether the
bullet cluster poses {\it any new challenges} to MOND at galaxy
cluster scales. It is indeed well-known that the dynamical mass 
from the X-ray temperature profiles of clusters in MOND exceeds their baryonic content (Aguirre et al. 2001). As a fix, a component of $\sim 1-2$eV neutrinos
has been invoked to explain cluster cores (Sanders 2003) and the Cosmic Microwave Background (McGaugh 2004). 

Nevertheless, applying a {\it new method} developed to fit the detailed weak lensing map of the bullet cluster, 
we place {\it robust constraints} on the dark matter density
in non-spherical non-equilibrium geometry. We test a range of gravity 
theories {\it interpolating smoothly between MOND and GR}.
We call these models the $\mu$-gravity Hot Dark Matter models 
($\mu$HDM, with appropriate interpolating function $\mu$ and hot neutrino content)
in contrast to its contending $\Lambda$CDM model 
(with appropriate cosmological constant $\Lambda$ and cold DM content).
Note both cosmologies can drive late-time acceleration of the universe
(Diaz-Rivera et al. 2006, Zhao 2006), and form structures and CMB anisotropies (Skordis et al. 2006, Dodelson \& Liguori 2006).


GR is a limiting case of the multi-field theory of gravity, TeVeS (Bekenstein
2004) as Newtonian gravity is a limiting case of MOND 
(Bekenstein \& Milgrom 1984, BM84 hereafter).  
In these theories, the total potential is due to the usual Newtonian 
potential of baryons plus a baryon-tracking scalar field (see e.g. Zhao \& Famaey 2006),
which creates the DM or MOND effect.  While there are subtle differences with MOND in non-spherical
geometries, in the limiting case of scale-free flattened models, AFZ06
showed that using the Poisson-like equation of BM84 
for the total gravitational potential $\Phi$ was a
reasonable approximation to the multi-field approach.

We thus investigate here a class of MOND-like non-linear laws of gravity, where 
the gravitational potential $\Phi$ satisfies the Poisson-like equation of BM84.
The average mass density $\overline{\rho}(<r)$ or the total mass (e.g., baryons and neutrinos) of the system enclosed inside any radius $r$ centered on {\it any position} will thus be
estimated from the divergence theorem with an effective gravitational parameter $G_{\rm eff}$ (not a constant):
\beq\label{Menclose} 
{M_{\rm bary}(r) + M_{\nu}(r) } 
= \int
{\partial \Phi(r,\theta,\psi) \over \partial_r}  { dA \over 4 \pi G_{\rm eff} }, 
\eeq where the surface area element $dA=\sin(\theta) d\theta d\psi$ 
and the interpolating function (cf. AFZ06)
${G \over G_{\rm eff}(x)} \equiv \mu(x) = 1 - \left[ {1+\alpha x \over 2} +
\sqrt{\left({1-\alpha x \over 2}\right)^2 + x }\right]^{-1},
$ where $x= {\left|\grad \Phi\right| \over a_0}$.
The case $\alpha \rightarrow \infty$ (i.e. $G_{\rm eff}=G$) corresponds to General
Relativity, whilst $\alpha=0$ is the toy-model (Eq.64) of Bekenstein (2004), 
and $\alpha=1$ is the simple $\mu$-function which has a better fit 
to the terminal velocity curve of the Milky Way (Famaey \& Binney 2005). 
We also examine the standard interpolating function 
$\mu(x)= {x \over \sqrt{1+x^2}}$ for comparison with other works.

\begin{table*}
\centering
\caption{Best fit parameters of the convergence map. Squared velocities in
$(1000\kms)^2$ and scale radii, p in kpc.}
\begin{tabular}{|cccc|cccc|}
\hline\hline
$v_{CM1}^2$ &$v_{CM2}^2$&$v_{XR1}^2$&$v_{XR2}^2$&$p_{CM1}$&$p_{CM2}$&$p_{XR1}$&$p_{XR2}$\\
2.84&1.45&0.38&0.17&227.4&155.4&62.6&33.4\\
\hline\hline
$X_{CM1}$ &$Y_{CM1}$&$X_{XR1}$&$Y_{XR1}$&$X_{CM2}$&$Y_{CM2}$&$X_{XR2}$&$Y_{XR2}$\\
-416.7&-173.1&-209.0&1.2&293.0&-2.7&147.5&3.6\\
\hline\hline
\end{tabular}
\end{table*}

{\it Fitting the convergence map of a multi-centred X-ray cluster:}
In GR, the convergence map allows us to immediately
derive the underlying projected density of matter. However, as shown
in AFZ06, the situation is different in MOND, where {\it what you see}
(in terms of convergence) is {\it not always what you get} (in terms of
density). For that reason, we use a potential-density approach
hereafter: we fit the convergence map using a parametric set of
potentials, and then use the best-fit potential to derive the
corresponding surface density for
various choices of the gravity's interpolating function $\mu$.

The bullet cluster is 4-centred, the centres being the
positions of the main cluster's Collisionless Matter 
(referred to as CM1, including its member galaxies) 
and X-ray gas components (XR1), and
the sub-cluster's Collisionless Matter (CM2) and X-ray gas components (XR2).  We chose to
model those 4 mass components as 4 spherical potentials: note
however that, in non-linear gravities, 
the 4 mass densities corresponding to those spherical potentials will {\it not} linearly add up, especially when $\mu$ is rapidly varying with position inside the system.

We thus write the lens-potential as a superposition of four
potentials pinpointed at four centres $\vec{r}_i$:
\beq 
\Phi(\vec{r}) = \sum_{i=1}^{4} v_i^2 \ln \sqrt{1 + {|\vec{r}-\vec{r}_i|^2  \over p_i^2 }}.
\eeq 
Each potential is fully described by two parameters, the
asymptotic circular velocity $v_i$ and the scale length $p_i$. 
In GR these
potentials correspond to cored isothermal density profiles. 
These potentials are similar to the cusped potentials of
AFZ06, which were found to have less good fit to
the convergence map.

Using Fig. 1b of C06 we set up a coordinate system for the bullet
cluster. The centers XR1, XR2 and CM2 lie, to a first approximation, along the RA direction, which we chose as
our $x$-axis. Our $z$-axis is along the line of sight. As suggested
by Markevitch et al. (2004) and C06, we chose the four centres of the potential to be exactly in the $x$-$y$
plane with their $(X,Y)$ coordinates chosen at the four observed peaks.
As a consequence, the potential of Eq.(2) has 8 parameters.

The
parametric convergence map in the $x-y$ plane is simply computed by
linear superposition of the individual contributions to the
convergence from the four spherical potentials (see AFZ06), the
convergence of each solely depending on its parameters $v_i$ and
$p_i$, and the rescaled radius $s_i \equiv p_i^{-1}\sqrt{(x-X_i)^2+(y-Y_i)^2}$:
\footnote{Erratum of AFZ06: correcting a typo, their Eq.25   
should read $\kappa(R) ={(|\vec{R}-\vec{R}_i|^2-2p_i^2)D_{\rm
eff}\theta(|\vec{R}-\vec{R} _i|) \over
2|\vec{R}-\vec{R}_i|(|\vec{R}-\vec{R}_i|^2-p_i^2)}+ {2p_i D_ {\rm
eff} v_i^2 \over c^2(|\vec{R}-\vec{R}_i|^2-p_i^2)}$ where the bending
angle $\theta$ is given by Eq.23 of AFZ06}.

\beq
\kappa(x,y)=\sum_{i=1}^{4}
{\pi v_i^2 D_{\rm eff} \over c^2 p_i} 
\left[(s_i^2+1)^{-{1 \over 2}} + (s_i^2+1)^{-{3 \over 2}}\right].
\eeq

We then reproduced the observed convergence map (Fig. 1b of C06) by least-squares fitting the asymptotic velocities $v_i$ and concentration parameters $p_i$ of each of the spherical potentials, using $n=233$ points from the $\kappa=0.16,0.23,0.3,0.37$ contours (always with a constant number of points per contour length). We also tried moving the 2-coordinate centres well within the errors of the brightest cluster galaxy. 

The goodness of fit of the model $\kappa_{{\rm model},i}$ to observations  $\kappa_{{\rm obs},i}$
is given by a characteristic variance 
$
{1 \over n}\sum_{i=1}^n \left({ \kappa_{{\rm obs},i}-\kappa_{{\rm model},i}}\right)^2.$
Due care was taken to maximize the resemblance to the X-ray
gas features with the centres as marked in Fig. 1. Another constraint was trying to
ensure a reasonable mass of X-ray gas to conform with the estimates of
C06 and Bradac et al. (2006, hereafter B06; see discussion section). 
The best-fit parameters (listed in Table 1) yield the convergence map
shown in Fig.1 upper panel, with a variance $\sim 0.1 \times 0.07^2$,  
which is acceptable since $0.07$ is 
both the typical observational error at individual points 
and the convergence spacing between two neighbouring contours.

\begin{figure*}
\vskip -5cm
\includegraphics[]{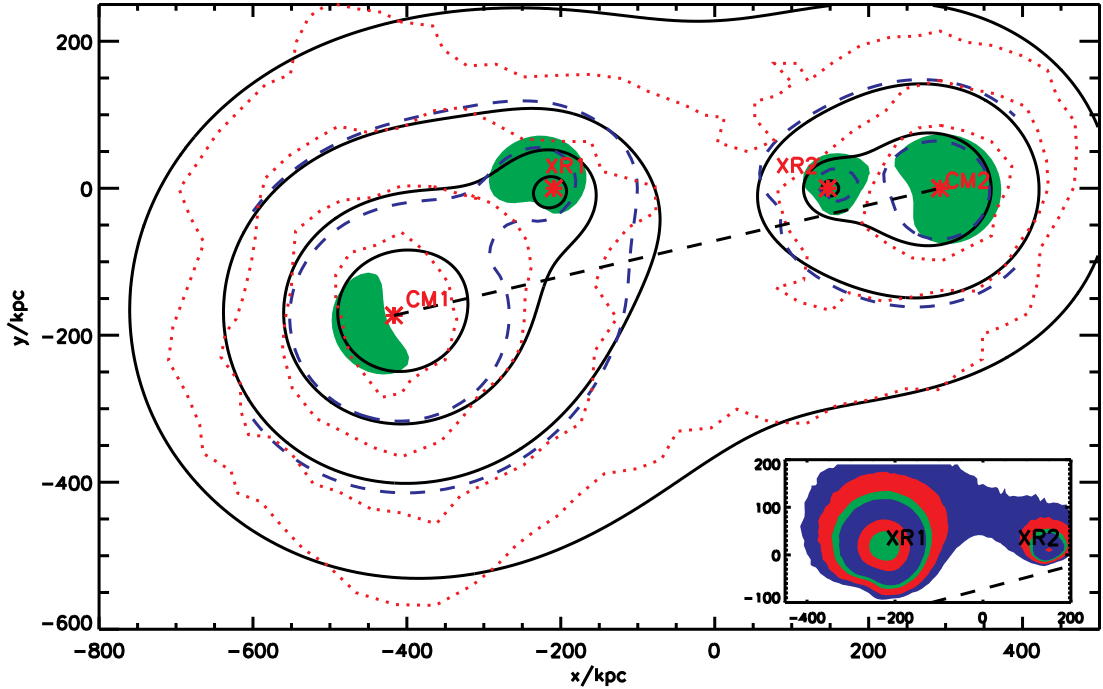}
\vskip -4.5cm
\caption{Our fitted convergence map (solid black lines) overplotted on the convergence map of C06 (dotted red lines) with x and y axes in kpc. The contours are from the outside
0.16,0.23,0.3 and 0.37. The centres of the four potentials we used are the red
stars which are labelled. Also overplotted (blue dashed line) are two contours of surface density [4.8 \& 7.2]$\times 10^2\msun\pc^{-2}$ for the MOND standard $\mu$ function; note slight distortions compared to the contours of $\kappa$.  The green shaded region
is where matter density is above $1.8\times 10^{-3}\msun\pc^{-3}$ and correspond to the clustering of 2eV neutrinos.
{\it Inset:} The surface density of the gas in the bullet cluster
predicted by our collisionless matter subtraction method for the standard
$\mu$-function. The contour levels are [30, 50, 80, 100, 200, 300]$ M_{\sun}pc^{-2}$. The origin in RA and dec is [$06^h58^m24.38^s$,-$55^o$56`.32]
}
\end{figure*}

\begin{table*}
\centering
\caption{Compares the results of C06 and B06 with estimates of projected mass in [$10^{13}M_{\sun}$] for gas around the X-ray centres and total mass around the lensing centres in three different gravities (GR, standard and simple $\mu$); a bias factor of $1.15-1.73$ should boost our {\it asymmetry-based} gas mass estimates here. 
B06 used an ellipse with semi-axes of 250~kpc and 150~kpc around XR1, which we compare with a circle of 180~kpc (the same area). 
The last column predicts the average matter densities in [$10^{-3} M_{\sun}pc^{-3}$] within 100kpc of the lensing peaks; the degeneracy of lensing to a stretching of the potential along the line of sight $\phi(x,y,z) \rightarrow \zeta \phi(x,y,z\zeta)$ means that
the central density predictions would be easily {\it lowered} if we adopted $\zeta < 1$.}
\begin{tabular}{|c|cc|cc|cc||}
\hline\hline
$\mu$ &$M^{gas}_{XR1}$&$M^{gas}_{XR2}$&$M_{CM1}$&$M_{CM2}$&$\overline{\rho_{CM1}}$&$\overline{\rho_{CM2}}$\\
\hline
&r$<$100/180kpc&r$<$100/80kpc&r$<$250kpc&r$<$250kpc&$r<100$&$r<100$\\
GR&1.05/1.97&0.33/0.27&21.7&17.2&2.63&2.59\\
standard $\mu$&0.97/1.79&0.29/0.24&9.0&6.78&2.26&2.34\\
simple $\mu$&0.74/1.33&0.21/0.18&7.13&6.42&1.66&1.76\\
C06/B06&0.66/2.0&0.58/0.42&20.0/28.0&21.0/23.0&&\\
\hline\hline
\end{tabular}
\end{table*}

{\it Masses of gas and collisionless matter in various postulated gravities:}
Applying Eq. (1) to our potential model we can predict
the matter volume density in the clusters, e.g., the values given in Table 2
up to a trivial degeneracy.
Integrating over the line of sight, we note that
the projected density contours are slightly different from that of convergence
contours in non-linear gravities (cf. dashed blue contours of Fig.1).  While
confirming AFZ06, this non-linear effect appears much milder than expected
earlier.

In order to match the observed X-ray gas mass,
which is a minor contributor to the lensing map, we use the asymmetry in the calculated surface
density to subtract off all the collisionless matter centred on the galaxies (CM1 and CM2). The key here is to notice the symmetry of galaxies 
around the dashed line joining the centres of the two galaxy clusters (cf. Fig.
1 upper panel). If we 
fold the map over the axis of symmetry subtracting the lower part from the 
upper part we are left with the majority of the gas. Then we
performed a straight forward numerical integration over the areas
defined above. The lower panel of Fig.1 demonstrates that this technique works well  
in separating the surface density of gas from the collisionless matter. 
The values for the gas mass for our three gravities
are given in Table 2.  Note this technique works less well for the sub cluster (XR2) as 
it lies quite close to the axis of symmetry and thus much gas is cancelled out by other gas. For GR only we can directly compare the gas corresponding to the potential and that
calculated by our subtraction method. For the main cluster, we find that integration of the surface density gives $2.32\times 10^{13} M_{\sun}$ within the 180kpc aperture which is 15\% more gas than estimated by asymmetry.  For the sub cluster we find $5.7\times 10^{12} M_{\sun}$ from integration within the 100kpc aperture of the gas center, 73\% more than from asymmetry.  As such, in MOND we can expect the gas
masses to increase by similar amounts and this helps to explain the low gas masses found, especially in the sub cluster (XR2). The reason our $\kappa$-map is skewed towards the gas peaks is a feature of the cored isothermal potentials. Table 2 shows we pack too much gas into the central 100kpc of the main cluster compared to that observed only for it to balance by 180kpc. Using a potential that correctly matches the gas density would not skew the map.

Table 2 also compares the B06 and C06 projected mass within a 250~kpc circular aperture 
centred on both galaxy clusters with our total
mass within these apertures for three gravities (GR, simple $\mu$ and standard $\mu$). Clearly, these amounts of mass 
exceed the observed baryons in gas and galaxies over the same apertures, 
by a factor of 3 even in MOND.
While very dense clumps of cold gas or MACHOs are still 
allowed by the missing baryon 
budget to reside in galaxy clusters without much collisions, 
we will focus on the possibilities of {\it fermionic particles}
being the unseen collisionless matter in the lensing peaks.

Following Tremaine \& Gunn (1979), we use the densest regions of the
collisionless matter to set limits on the mass of ordinary/sterile neutrinos.  A cluster core 
made of neutrinos of mass $m_\nu$ would have a maximum density (Sanders 2003)
satisfying ${\rho_{\nu}^{\rm max} \over 1.9\times 10^{-3}\msun\pc^{-3}}
\left({T \over 9\keV}\right)^{-3/2} = \left({m_\nu \over
2\eV}\right)^4$ where we adopted a mean temperature of 9 keV in the
two clusters.
Comparison with the regions of the highest volume density 
of matter shown in Fig.1 upper panel suggests 
that the relatively diffuse phase space
density in the bullet cluster is still consistent with ordinary $2\eV$
neutrinos making up the unseen collisionless component.
Note the lensing-predicted 3D density is generally non-unique 
due to the degeneracy of a line-of-sight stretching of the potential.
While a better fit to the gas mass and the lensing map could be 
produced by using several ellipsoidal potential components as opposed to 
the rigid four spheres with fixed centers here, 
the present model suffices as a demonstration.

{\it Discussion} The lensing-predicted 2 eV mass for neutrinos in the
non-equilibrium bullet cluster greatly strengthens the finding of Pointecouteau \&
Silk (2005) that ($m_\nu >1.6\eV$ using standard $\mu$) based on
spherical gas equilibrium of other clusters. 

Moreover, the 2eV neutrinos are falsifiable in the near future. 
At present it is compatible 
with {\it model-independent} experimental limits on electron neutrino mass
$m_{\nu,e}<2.2\eV$ from the Mainz/Troitsk experiments of counting
the highest energy $\beta$-decay electrons of ${\rm ^3H} \rightarrow
{\rm ^3He}^{+} + {\rm e}^{-} + \nu_e + 18.57 \keV$ (the more massive
the neutrinos, the lower the cutoff energy of electrons).  The
KATRIN experiment (under construction) will be able to falsify 2eV electron neutrinos at 95\% confidence within months of taking data in 2009.  Our prediction of a 4th (hot sterile) neutrino is fasifiable by the Booster Neutrino Experiment.

We note finally a couple of
discontinuities with the work of C06 and B06. Our
adopted lensing map of C06 (with a peak $\kappa=0.37$) 
implies a surface density that is too weak to form the observed large scale arcs 
in any gravity for sources at any redshift. 
Secondly, our MONDian models greatly reduce the amount of 
collisionless matter needed to fit the map of B06 in GR; 
a reduction by 3-4 times at 250 kpc. These masses are all integrated over the line of sight and as such give poor estimates of the mass {\it in the system} for comparison with the gravity independent gas masses.
Consistency between strong and weak lensing data remains to be understood
together with issues of smoothing, normalization and zero point
of the $\kappa$ maps.

Nonetheless, the data still convincingly require a dominant component of collisionless and most probably non-baryonic matter at cluster scales. 
%
A traditional misconception is that the
existence of a large quantity of non-baryonic matter would make a modified gravity theory such as MOND
{\it contrived or redundant}.  This is {\it not} the case with
ordinary hot neutrinos, which are too diffuse 
to either perturb the good MONDian fits to galaxy
rotation curves nor explain these curves in GR. 

As a tie-breaker between $\Lambda$CDM and $\mu$HDM we note that high-speed
encounters are rare in CDM simulations
(Hayashi \& White 2006).  The potential well of the main cluster would
be too shallow to accelerate the subcluster without
stronger-than-Newtonian gravity (Farrar \& Rosen 2006). Our MONDian
isothermal potential well (cf. Eq. 1 and Table 1) would accelerate the
two clusters to a maximum relative speed 
$v_{\rm max} = 
\sqrt{v_1^2\ln(1+(r/p_1)^2)+v_2^2\ln(1+(r/p_2)^2)} \sim 4800$km/s if at 
$\sim {r \over 0.33v_{\rm max}} \sim 1.8$
Gyrs ago the two clusters turned around from the Hubble flow at 
$r \sim 2 r_{\rm vir} \sim 2\times 1500$ kpc 
and free fall towards each other along the east-west direction.

\acknowledgements 
We acknowledge many insightful discussions on the implications and interpretation of our results with Stacy McGaugh and Douglas Clowe. We greatly appreciate the detailed reading and shrewd comments of the anonymous referee that strengthened the letter. GWA, HSZ and BF acknowledge hospitality from Beijing University and the NAOC. GWA acknowledges a PPARC studentship and an overseas fieldwork grant 04217, HSZ acknowledges support from a PPARC Advanced Fellowship and an Outstanding Overseas Young Scholarship from the Chinese Academy of Science. BF is an FNRS research associate, and acknowledges a grant from the DRI of ULB.



\end{document}